# Solving Poisson's Equation using Deep Learning in Particle Simulation of PN Junction

Zhongyang Zhang[1], Ling Zhang[1], Ze Sun[1], Nicholas Erickson[1], Ryan From[2], and Jun Fan[1]
[1] Missouri S&T EMC Laboratory, Rolla, MO, USA, zhongyang.zhang.hust@gmail.com, jfan@mst.edu
[2] Boeing Company, St. Louis, MO, USA

*Abstract*—Simulating the dynamic characteristics of a PN junction at the microscopic level requires solving the Poisson's equation at every time step. Solving at every time step is a necessary but time-consuming process when using the traditional finite difference (FDM) approach. Deep learning is a powerful technique to fit complex functions. In this work, deep learning is utilized to accelerate solving Poisson's equation in a PN junction. The role of the boundary condition is emphasized in the loss function to ensure a better fitting. The resulting I-V curve for the PN junction, using the deep learning solver presented in this work, shows a perfect match to the I-V curve obtained using the finite difference method, with the advantage of being 10 times faster at every time step.

## I. INTRODUCTION

Transient events like electrostatic discharge (ESD) and electromagnetic pulse (EMP) can dramatically affect the normal function of semiconductor devices, even causing permanent damage to the device in extreme cases. To investigate the macroscopic impact of transient events on devices, particle simulations at the microscopic level can be utilized to compute the dynamic motion of electrons and holes driven by fields created by external voltage sources [1]-[2]. To dynamically simulate particles, drift, diffusion, and scattering processes must be executed at each time step. Following these processes, the Poisson's equation also needs to be solved at the end of each time step in order to update the fields caused by the updated distribution of particles [3].

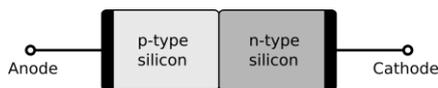

Figure 1. PN junction structure.

Utilizing the traditional finite difference method (FDM) to solve the Poisson's equation is fairly time-consuming, since each solving process may take a large number of iterations to converge. When there are numerous time steps in a full particle simulation, time consumption from this part will become considerable and even problematic. The popularity of deep learning in recent years provides a possible solution to accelerate this solving process due to its formidable function-fitting ability for complex problems.

With the development of big data and the improved high-performance computation ability brought by modern GPUs, deep learning is becoming an increasingly powerful tool which has achieved incredible success, especially in the area of image and speech recognition [4]-[5]. Recently, some researchers have utilized deep learning to deal with complex physical problems such as the quantum many-body problem [6], fluid dynamics[7]-[8], and to solve sophisticated equations like high-dimensional partial differential equations [9]-[10].

Some work has been done to accelerate the solving process of Poisson's equation in fluid simulations [8], which provides an idea to similarly speed up electromagnetic simulations. This was later demonstrated in [11]. In this work, the computational domain was represented by a 64 × 64 matrix, with Dirichlet boundary conditions. However, the boundary condition was avoided because a 32 × 32 inner area was used with the CNN, which significantly restricted its application since boundary condition is an important factor that needs to be considered as a variable in many scenarios. What's more, only the static electric potential distribution for one single step was predicted. In the particle simulation problem studied here, however, the input and the output matrix dimensions have to be the same, and the boundary condition of the potential distribution must be taken into account. Compared with single-step simulation, possible accumulative error may occur in multi-step particle dynamic simulation due to large amounts of simulation steps in series. This requires higher accuracy in deep learning.

This paper is organized as follows: In Section II the background of this particle simulation problem and traditional FD method to solve Poisson's equation will be introduced. In Section III, the implementation of the deep neural network will be described in detail, including input and output data, network structure, and the loss function. Calculation results using CNN solver will be shown in Section IV, while conclusions are drawn in Section V.

## II. PROBLEM DESCRIPTION

### A. Particle Simulation Problem

In this paper, a Monte Carlo simulator [12] is used to simulate the behavior of a PN junction under different bias conditions. The structure under simulation is shown in Figure 1. It consists of a p-region on the left and an n-region on the right, each with the same length (200 nm), height (80 nm) and doping density (1e23 $cm^{-3}$). The time step is set as 10 fs and the spatial mesh size is 10 nm. In this way, the computational domain is discretized, and both the potential and charge distributions are represented by the values at the edges using 9×41 matrices.

The flow of the Monte Carlo simulation is shown in Figure 2. It starts from initialization like the device configuration definition. Then the particles within the device are initialized by assigning their position, energy and wave vector randomly. After that, the charge distribution is updated for the first time, which is then fed into Poisson's solver to obtain the initial E field distribution. In the main loop, the calculated E-field drifts both electrons and holes [13]. Several scattering mechanisms are considered after each drift process, including polar optical scattering [14], non-polar optical scattering [1], acoustic scattering [14], and impurity scattering [15]. The number of particles moving into and out of the computational domain within each time step is counted to calculate the current through the two electrodes. At the end of each time step, the charge distribution is updated, which the Poisson solver then uses to update the potential distribution.

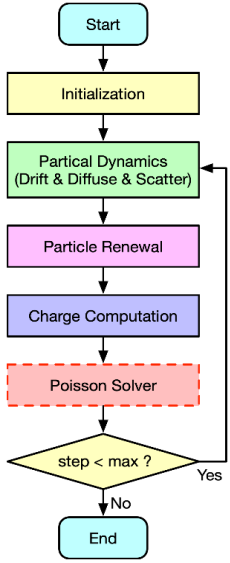

Figure 2. Flow of Monte Carlo simulation to understand device physics.

### B. Traditional Finite Difference Method

In order to simulate the behavior of the device precisely, the time step needs to be small enough. Since the Poisson solver is needed during each time step and the total particle simulation can include thousands of time steps, the solving process of the Poisson's equation needs to be fast and accurate.

The classical form of Poisson's equation can be written as

$$\nabla^2 V(r) = -\frac{\rho(r)}{\varepsilon} \quad (1)$$

where $\rho(r)$ is the charge density distribution and $V(r)$ is the potential distribution to be calculated. Since there is no closed-form solution for equation (1) for complex geometries, FDM is widely used to solve it [16].

The partial derivatives in the 2D case can be approximated through finite-difference, which is given as

$$\left(\frac{\partial^2}{\partial x^2} + \frac{\partial^2}{\partial y^2}\right)V(i,j) \approx \frac{V(i-1,j) + V(i+1,j)}{h^2} + \frac{V(i,j+1) - 4V(i,j) + V(i,j-1)}{h^2} \quad (2)$$

where $h$ is the length of each mesh cell. Finally, $V(i,j)$ can be solved as

$$V(i,j) = \frac{1}{4}[V(i-1,j) + V(i+1,j) + V(i,j-1) + V(i,j+1) + \frac{\rho(i,j)h^2}{\varepsilon}] \quad (3)$$

Therefore, equation (3) is used to update every voltage sample $V(i,j)$, using the voltages at the four neighboring grids and charge density $\rho(i,j)$ through many iterations, until the voltage values at all grid points converge. This traditional finite-difference method is straightforward to implement but not time-efficient, especially considering the numerous simulation steps needed in a whole particle simulation.

### III. CNN IMPLEMENTATION

#### A. Input and Output

The purpose of using a CNN is to predict the potential distribution based on the charge distribution and boundary conditions. Therefore, the CNN output is a potential matrix with the dimensions of 9 × 41, while the input has to contain a charge matrix with the same dimensions, which represents the distribution of electrons and holes. Further, the voltage bias is another factor that determines the boundary condition for the potential distribution. So another dimension was added in the input matrix to account for the boundary condition that must be satisfied, as shown in Figure 3. In the boundary condition matrix, the first and the last columns correspond to the boundary potential values specified on the two ends of the PN junction, while the rest of the values are all zeros.

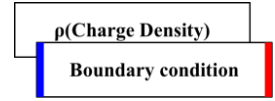

Figure 3. The input of the neural network, which has the shape of 2×9×41.

FDM was used to generate data for training and validation. Applied voltage bias was sampled every 0.1V between 0 to 2.8V, and for each voltage value, 20,000 simulation steps were executed and saved. Therefore, there are 580,000 pieces of data in total, among which 70% were selected as training data and 30% were used as validation data. Large amounts of data can prevent overfitting to some extent.

#### B. Network Structure

CNN is powerful network structure that can be used to extract complex and abstract features from images [4]. The input and output matrices described here can also be regarded as 'images', which makes them very suitable for CNN to deal with.

The CNN structure used in this work is shown in Figure 4. The input matrix has the dimensions of 512× 2 × 41 × 9, where 512 is the batch size and 2 × 41 × 9 are the dimensions of the combined charge and boundary condition matrix mentioned earlier. In the first convolutional layer, four parallel convolutional layers were used with a core size of 3×3, 5×5, 7×7, and 9×9, respectively, in order to capture features in different scales. Then these four feature maps were concatenated together and sent to several 3×3 convolutional layers in series to extract more abstract features.

The convolutional blocks in Figure 4 actually consist of several layers, as shown in Figure 5 for one of them as an example, including a convolutional layer, a batch normalization layer [17], a Leaky ReLU activation layer [18], and a Maxpool layer. The batch normalization layer is used after each convolutional layer and before each activation layer to optimize the performance of the network. The Leaky ReLU function was chosen because of its capability of preventing gradient vanishing problem in full range with fast computational speed and easy convergence. The Maxpool layer was used to reduce overfitting.

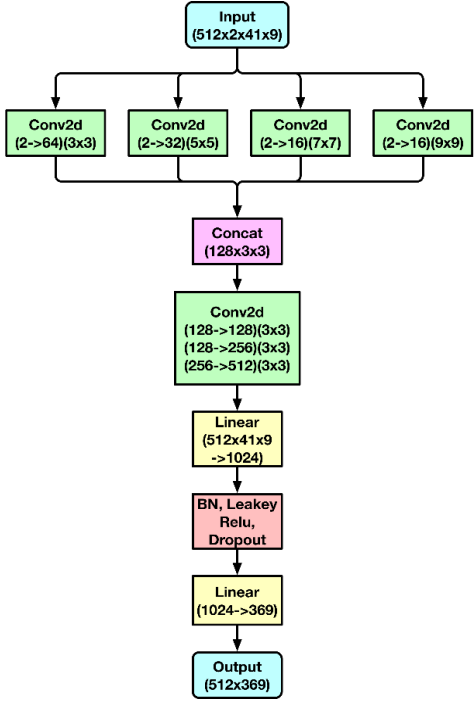

Figure 4. Convolutional neural network structure.

After the convolutional blocks, several linear fully-connected layers were used to reduce the data to the dimensions of the output matrix. Between the linear layers, batch normalization and Leaky ReLU layers were again applied. Afterward, a dropout layer [16] was used, which is also aimed at preventing overfitting. In summary, three methods were used to prevent overfitting and to improve the generalization performance of the network: 1. using more data for training; 2. using Maxpool layers; and, 3. using a dropout layer.

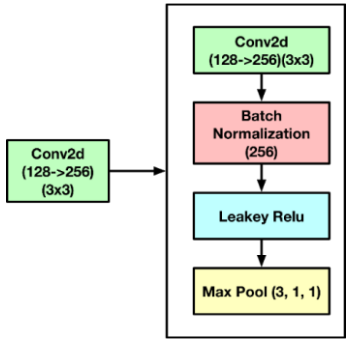

Figure 5. The inner structure of the Convolutional layer block.

## C. Loss Function

Since boundary conditions are important to the accuracy of the Poisson solver and hence that of the particle simulation, different to the loss function used in [8], one more term was added to emphasize the boundary condition and ensure a better prediction of boundary potential values, as shown by equation (6).

$$\text{Loss} = \sqrt{E((Y' - Y)^2)} + 2 \times \sqrt{E((B' - B)^2)} \quad (6)$$

where $E$ stands for mean value operator; Y' represents the predicted potential distribution; Y means the expected voltage distribution; and, B' means the predicted boundary-condition matrix while B is the expected boundary-condition matrix. The weights of these two terms are 1 and 2 respectively, which are determined empirically.

## IV. RESULTS

In the training, a learning rate of 0.0001 was used with Adam Optimizer. The change in training and validation loss can be found in Figure 6. After 16 epochs of training, the loss gradually converged and the training was stopped.

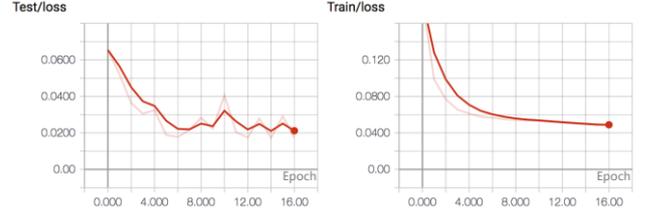

Figure 6. Testing and training loss graph of the proposed neural network.

In order to visualize the prediction results of the trained neural network, several random cases have been selected from the validation data, and the expected and the predicted potential distributions are compared in Figure 7.

It is found that the predicted potential distribution not only tracks the tendency well but also fits the boundary values perfectly. What's more, the predicted potential distribution is even smoother than the expected potential distribution calculated by FDM in some cases. The possible reason for the 'ripples' in the FDM results may be due to insufficient convergence criterion or spatial resolution. However, since the CNN model has been trained with plenty of data, it can probably learn that those 'ripples' are just noise that should be removed in the prediction results.

Now the trained CNN model is shown to be able to predict potential distribution for a single step. In order to further check the long-term performance to evaluate whether the accumulative error may become significant in particle simulation, in this work, the I-V curve of the PN junction was calculated using the CNN model. The CNN solver was integrated into the PN junction particle simulator to replace the original FDM solver and was ran for 20,000 simulation steps for each voltage bias (sampled every 0.05V from 0 to 2.8V). The current for each voltage bias was calculated by the CNN. Eventually, an I-V curve was obtained, as shown in Figure 8. It is clear that the I-V curve generated using the CNN solver agrees very well with the I-V curve calculated from the FD method. However, the CNN solver is much faster for solving each step. It takes only 3 milliseconds using NVIDIA GeForce GTX 1080 Ti Graphics Cards for every single step, while traditional FD method takes up to 40 milliseconds, which means the CNN solver can improve the solving speed of every single step by more than 10 times. This efficiency improvement is even more significant for larger numbers of simulation steps.

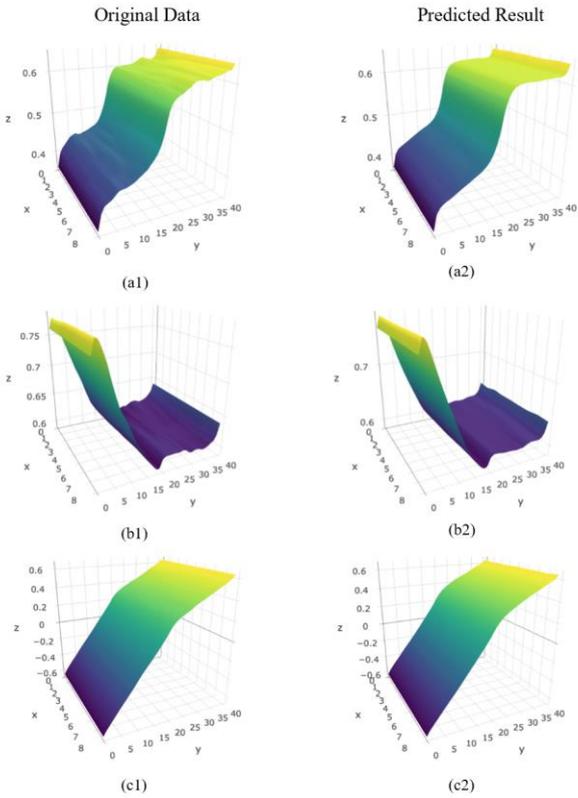

Figure 7. Comparisons of the simulated potential distributions using FDM (left) and the predicted potential distributions using the proposed CNN (right). Three samples (a,b,c) are included.

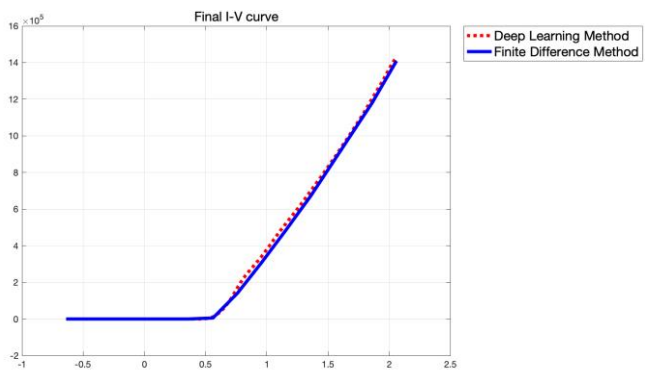

Figure 8. Comparison of the I-V curves generated by the deep learning solver (red dotted) and the traditional FD method solver (blue solid).

## V. CONCLUSION

This work has shown that deep learning techniques can be used to solve the Poisson's equation for dynamic particle simulation inside a PN junction. In this work, the boundary condition was emphasized in a hybrid loss function, and a neural network structure was carefully designed. The predicted potential distributions in the PN junction using the proposed CNN match very well with those simulated with traditional FD method. Also, the particle simulation using a deep learning solver at every time step resulted in an accurate I-V curve with very little accumulative error, while the calculation speed for solving Poisson's equation was improved by more than 10 times. In future work, a deep learning solver can be used to simulate transient voltage excitation in more complicated devices.

## VI. REFERENCES

[1] R. W. Hockney and J. W. Eastwood, *Computer simulation using particles*. crc Press, 1988.
[2] R. W. Hockney, R. A. Warriner, and M. Reiser, "Two-dimensional particle models in semiconductor-device analysis," *Electron. Lett.*, vol. 10, no. 23, pp. 484–486, 1974.
[3] K. Tomizawa, "Numerical Simulation of Submicron Semiconductor Devices (Artech House Materials Science Library)," *Boston MA USA Artech House*, 1993.
[4] A. Krizhevsky, I. Sutskever, and G. E. Hinton, "Imagenet classification with deep convolutional neural networks," in *Advances in neural information processing systems*, 2012, pp. 1097–1105.
[5] G. Hinton *et al.*, "Deep neural networks for acoustic modeling in speech recognition: The shared views of four research groups," *IEEE Signal Process. Mag.*, vol. 29, no. 6, pp. 82–97, 2012.
[6] D. Vasileska, S. M. Goodnick, and G. Klimeck, *Computational Electronics: semiclassical and quantum device modeling and simulation*. CRC press, 2016.
[7] J. N. Kutz, "Deep learning in fluid dynamics," *J. Fluid Mech.*, vol. 814, pp. 1–4, 2017.
[8] J. Tompson, K. Schlachter, P. Sprechmann, and K. Perlin, "Accelerating eulerian fluid simulation with convolutional networks," *ArXiv Prepr. ArXiv160703597*, 2016.
[9] N. Srivastava, G. Hinton, A. Krizhevsky, I. Sutskever, and R. Salakhutdinov, "Dropout: a simple way to prevent neural networks from overfitting," *J. Mach. Learn. Res.*, vol. 15, no. 1, pp. 1929–1958, 2014.
[10] J. Han and A. Jentzen, "Overcoming the curse of dimensionality: Solving high-dimensional partial differential equations using deep learning," *ArXiv Prepr. ArXiv170702568*, 2017.
[11] W. Tang *et al.*, "Study on a Poisson's equation solver based on deep learning technique," in *2017 IEEE Electrical Design of Advanced Packaging and Systems Symposium (EDAPS)*, 2017, pp. 1–3.
[12] C. Jacoboni and P. Lugli, *The Monte Carlo method for semiconductor device simulation*. Springer Science & Business Media, 2012.
[13] N. Nintunze and M. A. Osman, "Hole drift velocity in the warped band model of GaAs," *Semicond. Sci. Technol.*, vol. 10, no. 1, p. 11, 1995.
[14] T. Brudevoll, T. A. Fjeldly, J. Baek, and M. S. Shur, "Scattering rates for holes near the valence-band edge in semiconductors," *J. Appl. Phys.*, vol. 67, no. 12, pp. 7373–7382, 1990.
[15] E. Conwell and V. F. Weisskopf, "Theory of impurity scattering in semiconductors," *Phys. Rev.*, vol. 77, no. 3, p. 388, 1950.
[16] J. R. Nagel, "Solving the generalized poisson equation using the finite-difference method (fdm)," *Lect. Notes Dept Electr. Comput. Eng. Univ. Utah*, 2011.
[17] S. Ioffe and C. Szegedy, "Batch normalization: Accelerating deep network training by reducing internal covariate shift," *ArXiv Prepr. ArXiv150203167*, 2015.
[18] B. Xu, N. Wang, T. Chen, and M. Li, "Empirical evaluation of rectified activations in convolutional network," *ArXiv Prepr. ArXiv150500853*, 2015.